\documentclass[a4paper,12pt]{article}
\title{Comment on "Solution of the problem of catastrophic relaxation of
homogeneous spin precession in superfluid $^3$He-B".}
\author {I.A.Fomin\\
P. L. Kapitza Institute for Physical Problems, \\
ul. Kosygina 2, 119334 Moscow,Russia}
\date{ }
\begin{document}
\maketitle
\begin{abstract}
The argument of Bunkov, L`vov and Volovik contains errors, making
the obtained result non-convincing.

\end{abstract}

The paper of Bunkov, L`vov and Volovik (BLV, following their own
abbreviation) \cite{BLV} is the second paper, published recently
in "Pis`ma v ZhETF" and addressing the problem of theoretical
explanation of "catastrophic relaxation" in superfluid $^3$He-B.
It differs from the paper \cite{SF}, published before in setting
of the problem. In Ref. \cite{SF} a problem of stability of
spatially uniform precession of spin in $^3$He-B with the order
parameter following motion of spin in the Brinkmann and Smith
configuration is analyzed. It is shown in this  paper, that the
described precession at sufficiently low temperatures experiences
Suhl instability, i.e. decays in parametrically excited spin
waves. Coupling between the uniform precession and spin waves is
provided by the anisotropy of spin-wave velocity. Brinkmann and
Smith stationary solution of Leggett equations serves as a basis
for interpretation of pulsed NMR experiments in $^3$He-B. For
other configurations precession of spin is stationary only on the
average, in the limit $\Omega/\omega_L\ll 1$, where $\Omega$ is
the Leggett frequency and $\omega_L$ - the Larmor frequency. Even
in this limit for most of configurations, including those,
considered by BLV, the uniform precession is unstable  \cite{fom1}
due to the same mechanism, which is responsible for the
instability of precession in the A-phase.  So, the result of Ref.
\cite{SF} means that Suhl instability sets  a limit in temperature
for possibility to realize precession of spin in $^3$He-B even in
the ideally uniform conditions. Only this limiting temperature can
be taken as a physically meaningful definition of $T_{cat}$.

In Ref. \cite{BLV} a different question is addressed: why the
stationary precession of spin can not be observed below certain
threshold temperature in the particular conditions of experiments
Refs. \cite{nyeki,bun1}, which are considered as typical? Figure
in Ref. \cite{BLV}  shows that at these conditions the walls of
container perturb Brinkman and Smith configuration in a
significant part of the container. BLV argue, that in the
"typical" situation the idealization of Ref. \cite{SF} does not
apply  and suggest instead their own treatment of the problem also
based on  Suhl instability and on a locally uniform precession,
but out of Brinkmann and Smith configuration. The essential point
in the argument is that for configurations different from that of
Brinkmann and Smith there exist additional coupling of a uniform
precession to spin waves, originating from the dipole energy and
this coupling can be dominant in "typical" experimental
conditions.

There is no doubt that analysis of validity of an idealization for
a particular experimental set-up is important part of
interpretation of experimental data and that a proper
generalization of theory in a way, that makes possible its
application to a wider class of experimental conditions is very
useful. Unfortunately this can not be said about the
generalization, suggested in Ref. \cite{BLV}. The argument of BLV
contains ambiguous assumptions and non justified approximations.
The most disturbing of these are the following:

1. The increment $V_{BLV}\sim\Omega^2/\omega_L$. BLV argue that
for small anisotropy of spin-wave velocity $\mu$ there exist a
region of magnetic fields $\mu\omega_L\ll\Omega\ll\omega_L$, where
this increment is greater then that, originating from the
anisotropy itself. But they do not take into account the existence
of another instability \cite{fom1}, which was mentioned above.
This instability  also has increment of the order of
$\Omega^2/\omega_L$. Why this mechanism can be disregarded?

2. For comparison with  experiments BLV average the found
increment over the volume of the cell. According to the figure in
Ref. \cite{BLV} variation of the increment $V$ is not small in
comparison with the average increment $\langle V\rangle$. In this
situation substitution of $\exp(\langle V\rangle t)$ instead of
$\langle\exp(Vt)\rangle$ is not correct operation - two ways of
averaging can give very different results.

3. The numerical example of Ref. \cite{BLV} which represents the
"typical" experimental conditions with $\Omega$ = 244 kHz and
$\omega_L$ = 460 kHz demonstrates a good agreement of the
calculated within the BLV scheme temperature of catastrophic
relaxation $T_{cat}$ = 0.5$T_c$ with the measured $T_{cat}$ =
0.47$T_c$. The square of wave vector of the excited spin waves
enters equation (28) of Ref. \cite{BLV} from which $T_{cat}$ is
evaluated. In this evaluation BLV use the linear dispersion law
for the $\gamma$-mode. When the dipole energy is taken into
account the $\gamma$-mode acquires a gap
$\omega_s(k)=\sqrt{[\Omega(s,l)]^2+(ck)^2}$. The value of
$\Omega(s,l)$ depends on $s$ and $l$ varying within the container
as shown on the figure in Ref. \cite{BLV}. The most important is a
region near the maximum of the increment. Consider as
representative the point, where $s=l\approx\cos28^o$. For this
point $\Omega(s,l)\approx 0.90\Omega\approx$ 220 kHz, i.e.
$\Omega(s,l)$ is only slightly smaller then $\omega_L/2$ = 230
kHz. With these parameters conservation of energy for the gapped
dispersion law requires much smaller values of $k$ then for the
linear dispersion. In the considered example the proper account of
the gap brings $k^2$ ten times down i.e. renders additional factor
of $\approx$10 in the r.h.s of Eq.(28). This factor definitely
must have a damaging effect on the obtained agreement. It is
important also, that for small $k$ breaks down the local
approximation, used by BLV.

In view of the above remarks the statement made in the conclusion
of Ref. \cite{BLV}:  "Our analytical result for the onset of the
parametric instability due to this mechanism is in a good
quantitative agreement with experimental results" does not
represent the situation correctly. Consequently, the problem at
which BLV are aiming is not solved in their paper.

On the other hand the message of BLV that conditions in the
experiments \cite{nyeki,bun1}  substantially deviate from the
ideal conditions assumed in Ref. \cite{SF} is a serious warning.
It means that for experimental investigation of the "intrinsic"
mechanism of catastrophic relaxation, suggested in Ref. \cite{SF},
larger containers and stronger magnetic fields have to be used,
like in the experiments Ref. \cite{lee}.


\begin{thebibliography}{99}


\bibitem{BLV} Yu. M. Bunkov, V. S. L`vov, G. E. Volovik, Pis`ma v ZhETF, {\bf 83}, 624
(2006), cond-mat/0605386.

\bibitem{SF} E. V. Surovtsev, I. A. Fomin, Pis`ma v ZhETF, {\bf 83}, 479 (2006).

\bibitem{fom1} I. A. Fomin, ZhETF {\bf 78}, 2392 (1980) [JETP,
{\bf 51}, 1203 (1980)] .

\bibitem{nyeki}  Yu. M. Bunkov, V.V. Dmitriev, Yu.M. Mukharsky et al.,
{\it Europhysics Lett.} {\bf 8}, 645 (1989).

\bibitem{bun1}  Yu. M. Bunkov , J. Low Temp. Phys., {\bf 135},337 (2004).

\bibitem{lee} D. A. Geller and D. M. Lee,  Phys. Rev. Lett.,
               {\bf 85}, 1302 (2000)



\end{thebibliography}
\end{document}